\documentclass[twocolumn,tightenlines,floats,floatfix,prc,nofootinbib,preprintnumbers]{revtex4}

\usepackage{hyperref}
\usepackage{enumitem}

\def\bea{\begin{eqnarray}}
\def\eea{\end{eqnarray}}

\def\st#1{{\kern-4pt} \not\!#1}

\def\sp{\kern +3pt}
\def\sm{\kern -3pt}

\usepackage[usenames]{color}

\def\be{\begin{equation}}
\def\ee{\end{equation}}
\def\ba{\begin{eqnarray}}
\def\ea{\end{eqnarray}}

\usepackage{graphics}
\usepackage{graphicx}
\usepackage{epsf} 
\usepackage{amsmath}
\usepackage{amssymb}
\usepackage{slashed}
\usepackage{bm}

\begin{document}

\phantom{0}
\vspace{-0.2in}
\hspace{5.5in}

\preprint{ {\bf LFTC-18-8/29}}

\vspace{-1in}

\title
{\bf Covariant quark model for the electromagnetic 
structure of light nucleon resonances}
\author{G.~Ramalho
\vspace{-0.1in} }


\affiliation{
Laborat\'orio de F\'{i}sica Te\'orica e Computacional -- LFTC,
Universidade Cruzeiro do Sul, 01506-000, S\~ao Paulo, SP, Brazil}

\vspace{0.2in}
\date{\today}

\phantom{0}

\begin{abstract}
We present estimates from the covariant spectator quark model 
for the electromagnetic transition form factors for the resonances
$N(1440)\frac{1}{2}^+$, $N(1535)\frac{1}{2}^-$,
$N(1520)\frac{3}{2}^-$ and $\Delta(1232)\frac{3}{2}^+$,
at intermediate and large square momentum transfer ($Q^2$).
The calculations associated to the 
$N(1440)\frac{1}{2}^+$, $N(1535)\frac{1}{2}^-$ and 
$N(1520)\frac{3}{2}^-$ states are based exclusively  
on the parametrizations derived for the nucleon.
In the case of the $\Delta(1232)\frac{3}{2}^+$ (isospin 3/2),
we use lattice QCD data to estimate the radial structure,
and take into account the well known effects associated with 
the pion cloud dressing of the baryons.
Our estimates are based mainly on the valence quark degrees 
of freedom and are in good agreement with the data 
for $Q^2 > 2$ GeV$^2$,
with a few exceptions.
The present predictions can be tested in a near future 
for large $Q^2$ at the Jefferson Lab -- 12 GeV upgrade.    
\end{abstract}

\vspace*{0.9in}  
\maketitle

\section*{BACKGROUND}

With the construction of the modern accelerators 
a significant amount of data associated 
with the electromagnetic structure of the nucleon ($N$)
and the nucleon resonances ($N^\ast$) 
has been collected, at intermediate and large square momentum transfer ($Q^2$).
The new data, parametrized in terms of 
$\gamma^\ast N \to N^\ast$ transition form factors, 
call for the development of relativistic theoretical models 
that can be applied to the description of the present data,  
as well as the expected results 
from the Jefferson Lab --12 GeV upgrade~\cite{NSTAR,SRap}.

In the large-$Q^2$ region, the $\gamma^\ast N \to N^\ast$ transitions 
are expected that to be dominated by the valence quark degrees of freedom.
One of the quark models that includes relativity 
is the covariant spectator quark model~\cite{Nucleon,SRap}.
The model was originally developed to study the electromagnetic 
structure of the nucleon~\cite{Nucleon}.
The motivation for the model was to test if the 
new results from Jefferson Lab (JLab)~\cite{Jones00} 
for the ratio between the electric and 
the magnetic proton form factors, could be explained 
by a simple quark model based on a three-constituent quark structure
(quarks with their own internal structure).
The JLab experiments have shown that the proton  
electric and magnetic form factors 
have different dependences on $Q^2$, 
suggesting  a difference 
between the electric charge and the magnetic dipole 
density distributions~\cite{Nucleon,Jones00}.

The model has three basic ingredients: 
\setlist{nolistsep}
\begin{enumerate}[label=(\roman*)] 
\itemsep0em 
\item
the wave function of the baryon (including the nucleon) 
can be represented in terms of the spin-isospin structure of 
the individual quarks based on 
the $SU_S(2) \times SU_F(3)$ spin-flavor symmetry,
rearranged as an active quark and a spectator quark-pair~\cite{Nucleon,Omega};
\item
the three-quark system  can be reduced  to a quark-diquark system, 
parametrized by a radial wave function $\psi_B$,
integrating into the quark-pair degrees of freedom~\cite{Nucleon,Nucleon2,Omega}; 
\item
the electromagnetic structure of the quark is 
parametrized by quark isoscalar/isovector form factors $f_{i \pm}(Q^2)$
($i=1$ for Dirac, and $i=2$ for Pauli),
which simulate the substructure associated with the gluons 
and quark-antiquark effects, 
and it is parametrized using 
the vector meson mechanism~\cite{Omega,LatticeD,Lattice}.
\end{enumerate}

A very good description of the new JLab results 
as well as the neutron data
is obtained when we calibrate  
the two building blocks 
 of the model: the quark form factors and 
the nucleon radial wave function, $\psi_N$,
by the proton and neutron form factor data~\cite{Nucleon}.
The model can then be extended to nucleon resonances,
in particular to the lightest $N^\ast$ states $J^P$
(spin $J$ and parity $P$).  


\section*{RESULTS}

\begin{figure*}[t]
\vspace{.6cm}
\centerline{\mbox{
\includegraphics[width=2.2in]{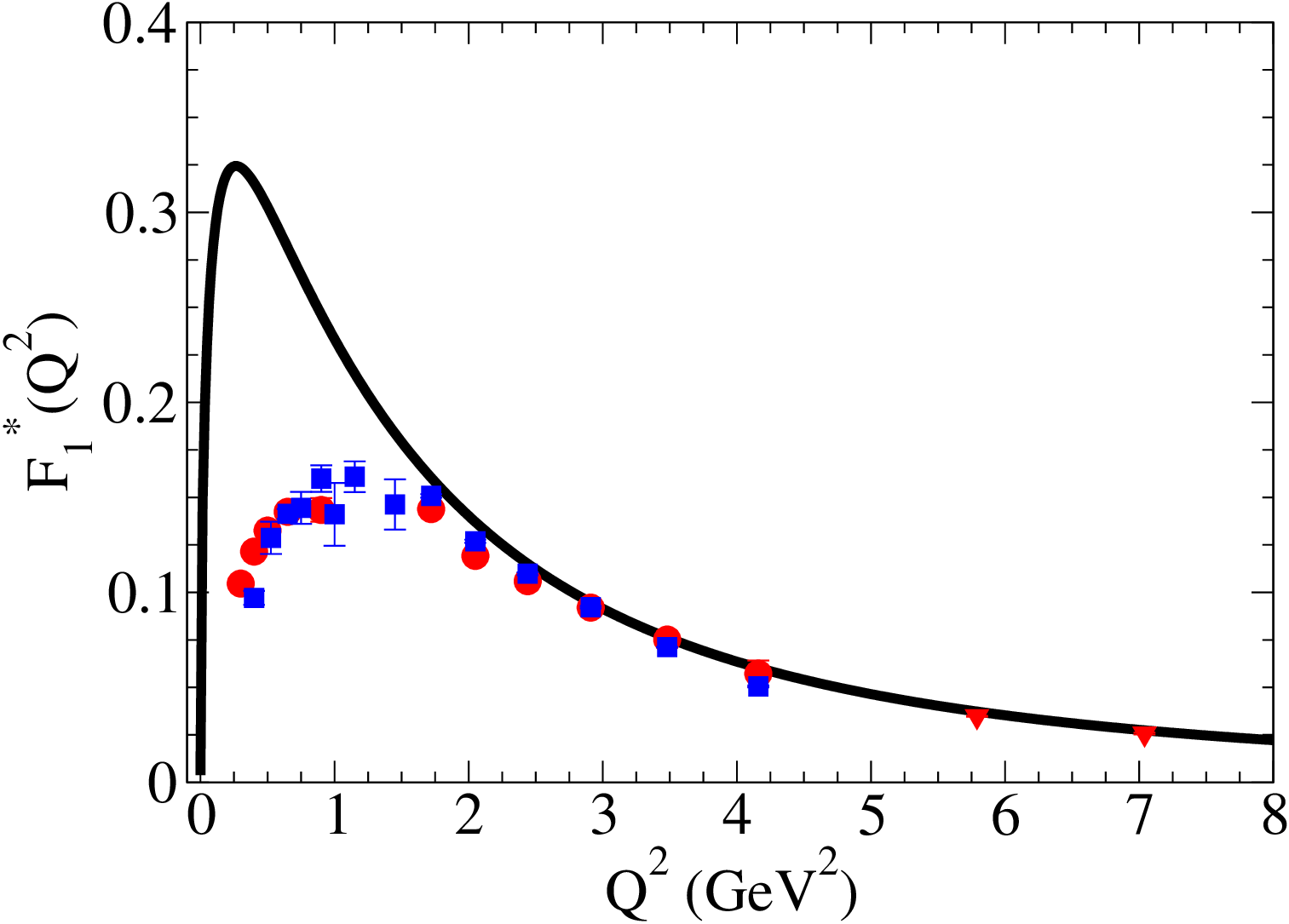}  \hspace{.2cm}
\includegraphics[width=2.2in]{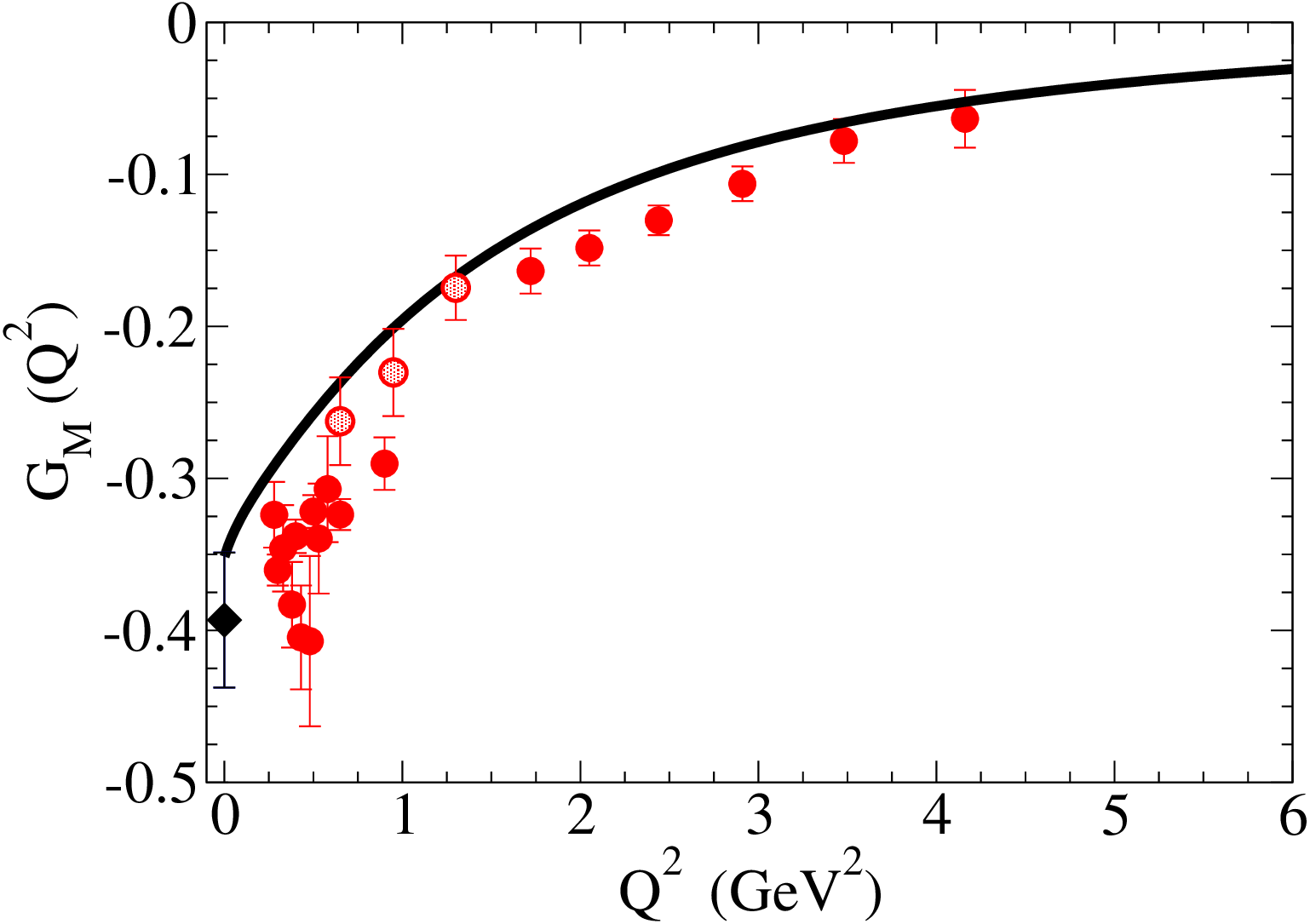}  \hspace{.2cm}
\includegraphics[width=2.2in]{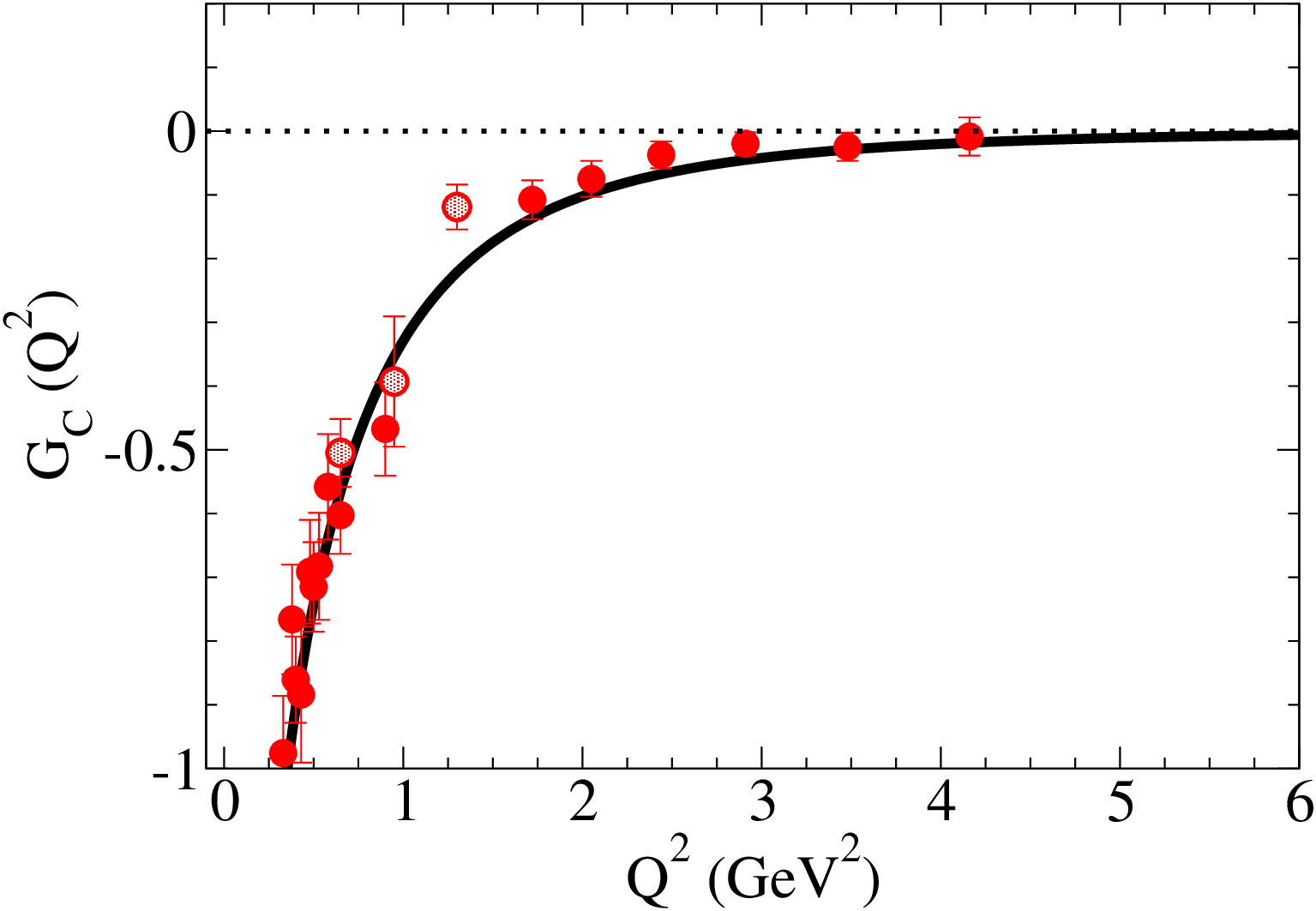}}}
\caption{\footnotesize
{\bf Left:}
$\gamma^\ast N \to N(1535)$ Dirac transition form factor.
{\bf Center and right:} 
$\gamma^\ast N \to N(1520)$ transition form factors:
Magnetic dipole (center) and Coulomb quadrupole (right).
Data from JLab (circles)~\cite{MokeevDatabase} 
and MAID (squares)~\cite{MAID}.
}
\label{figNegativeP}
\end{figure*}

The covariant spectator quark model has been 
in the recent years extended 
to the negative parity states $N(1535)\frac{1}{2}^-$
and $N(1520)\frac{3}{2}^-$~\cite{N1535,N1520,SQTM}.
The analytic expressions for the transition form factors
became simpler when we consider the semirelativistic approximation~\cite{SRap}.
In that approximation the radial wave function of the resonance ($\psi_R$)
has the same form as the wave function of the nucleon ($\psi_N$) and 
the difference of masses between the nucleon ($M$) 
and the resonance ($M_R$) is neglected in the overlap of the radial functions.
In these conditions the transition form factors 
can be determined without any additional assumption.
The only input are the parametrization of the quark form factors 
and the nucleon radial wave function, 
both determined in the study of the nucleon~\cite{Nucleon}.
The results for the $\gamma^\ast N \to N(1535)$ and   $\gamma^\ast N \to N(1520)$
form factors are presented in Fig.~\ref{figNegativeP},
in comparison with the data from JLab~\cite{MokeevDatabase}
and MAID~\cite{MAID}.

\begin{figure*}[t]
\vspace{.6cm}
\centerline{\mbox{
\includegraphics[width=3.1in]{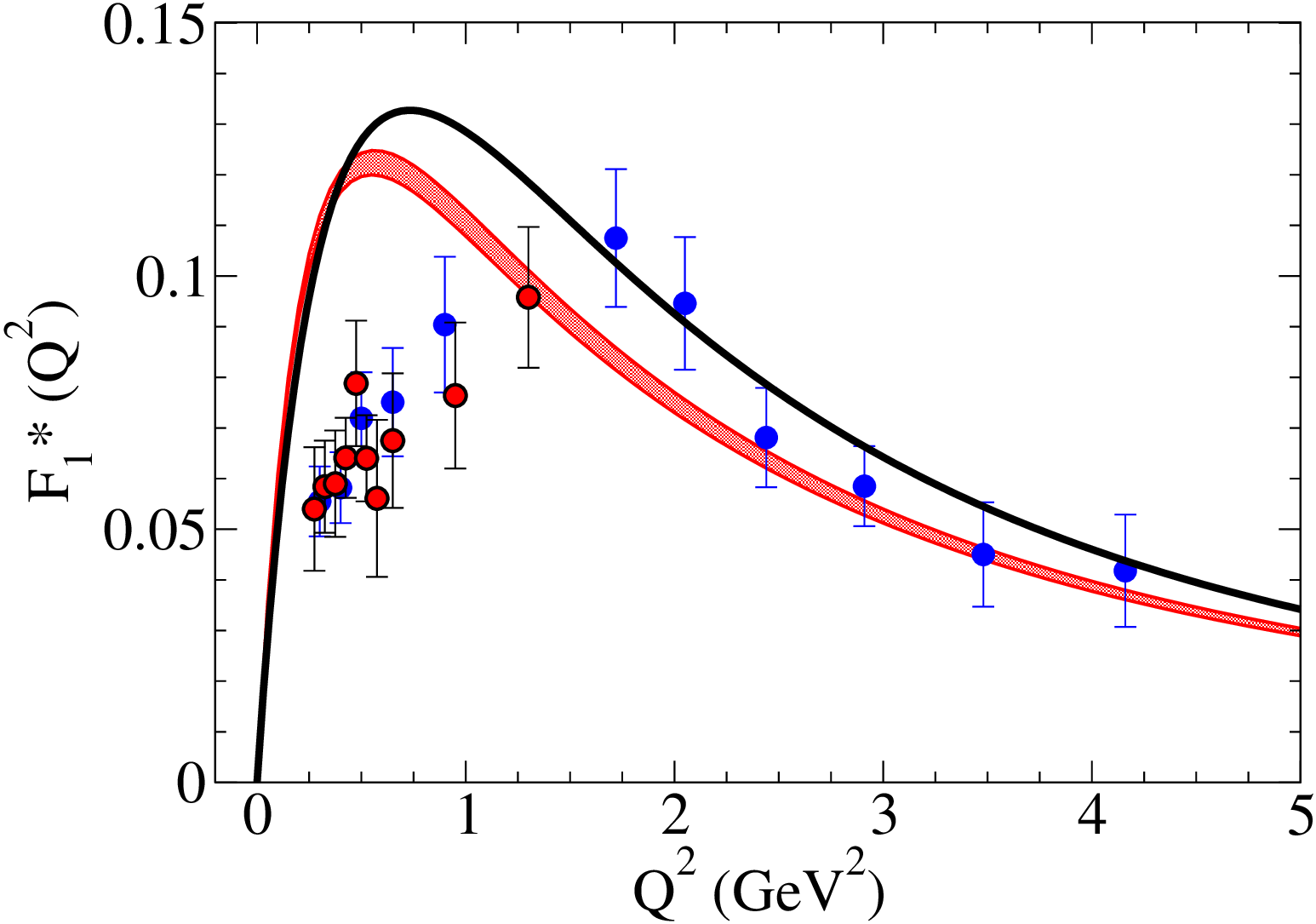} \hspace{.5cm}
\includegraphics[width=3.1in]{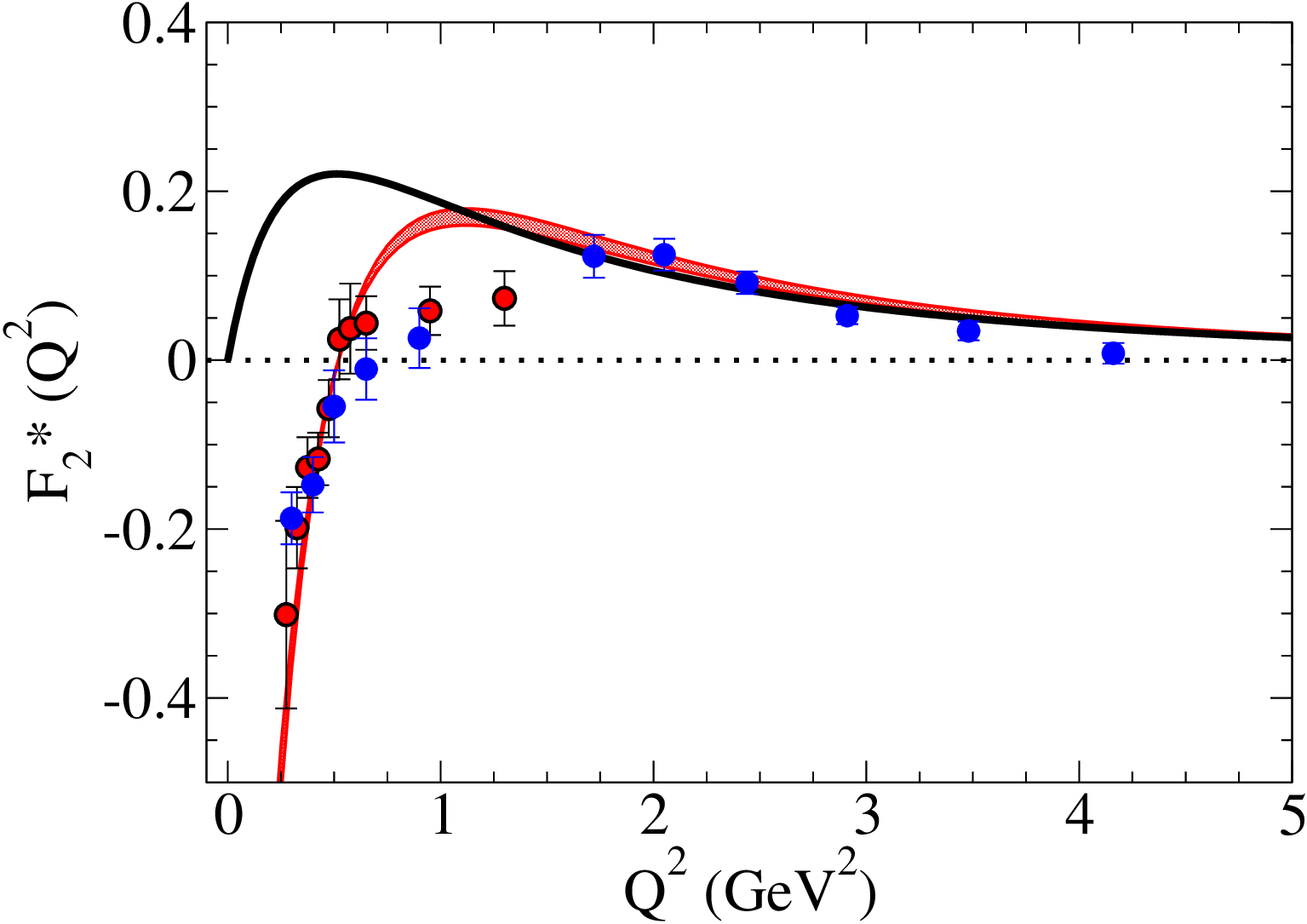}
}}
\caption{\footnotesize
$\gamma^\ast N \to N(1440)$ transition form factors.
The results from the covariant spectator quark model 
are represented by the solid line.
The results from the holographic model from Ref.~\cite{Roper-AdS}
are represented by the red band.
Data from JLab~\cite{MokeevDatabase}.}
\label{figRoper}
\end{figure*}

In general, we obtain a good overall description 
of the data, particularly for $Q^2 > 2$ GeV$^2$.
The exceptions are the Pauli form factor $F_2^\ast$ in the case of $N(1535)$ 
and the electric form factor $G_E$ in the  case of $N(1520)$ at low $Q^2$  
(not shown here)~\cite{SRap}. 
These deviations may be interpreted as an indication 
that the meson cloud effects,
not included in the present framework, 
may be significant at low $Q^2$.
For $N(1535)$ it was shown that the calculations 
from valence quark 
contributions and meson cloud contributions 
to $F_2^\ast$ have different signs, which may lead 
to significant cancellation between 
those effects~\cite{N1535-2,Jido,Jido08,SQTM}. 
The consequence of that cancellation is the 
correlation between the transverse and scalar amplitudes 
for large $Q^2$: 
\ba
S_{1/2} = - \frac{\sqrt{1 + \tau}}{\sqrt{2}} \frac{M_R^2- M^2}{2  M_R Q} A_{1/2},
\ea 
where $\tau= \frac{Q^2}{(M + M_R)^2}$.
The previous relation agrees remarkably well with 
the available data~\cite{N1535-2}.
In the case of $N(1520)$, the model fails at small $Q^2$, for $G_E$, 
because it predicts that the longitudinal amplitude 
$A_{3/2} \propto (G_E + G_M)$ vanishes,
contrary to the experimental evidences
($A_{3/2} \ne 0$)~\cite{NSTAR,MokeevDatabase}. 
This discrepancy can be understood admitting that $A_{3/2}$ 
is dominated by meson cloud effects, as discussed in detail 
in Refs.~\cite{N1520,SQTM,SRap,N1520-2}.

The model has also been  applied to the description of 
the $N(1440)\frac{1}{2}^+$~\cite{Roper,Roper2}, 
traditionally interpreted 
as the first radial excitation of the nucleon~\cite{NSTAR,Aznauryan07}.
In this case the radial wave function of the resonance 
is written in the form  
$\psi_R (\varkappa) = g(\varkappa) \psi_N(\varkappa)$
where $\varkappa$ is a variable defined in terms of the quark-diquark 
relative momentum~\cite{Nucleon,Roper,Roper2}.
The function $g(\varkappa)$ is expressed 
in a form compatible with the expected asymptotic behavior 
for the transition form factors at large $Q^2$ 
and includes one adjustable parameter.
This parameter is determined by the condition 
that the nucleon and the $N(1440)$ are orthogonal states.
Once defined  $\psi_R (\varkappa)$, the transition form 
factors are determined without the inclusion of any extra parameters,
except for the ones included in the  parametrization for 
$\psi_N$~\cite{Nucleon}.

\begin{figure*}[t]
\vspace{.6cm}
\centerline{\mbox{
\includegraphics[width=3.1in]{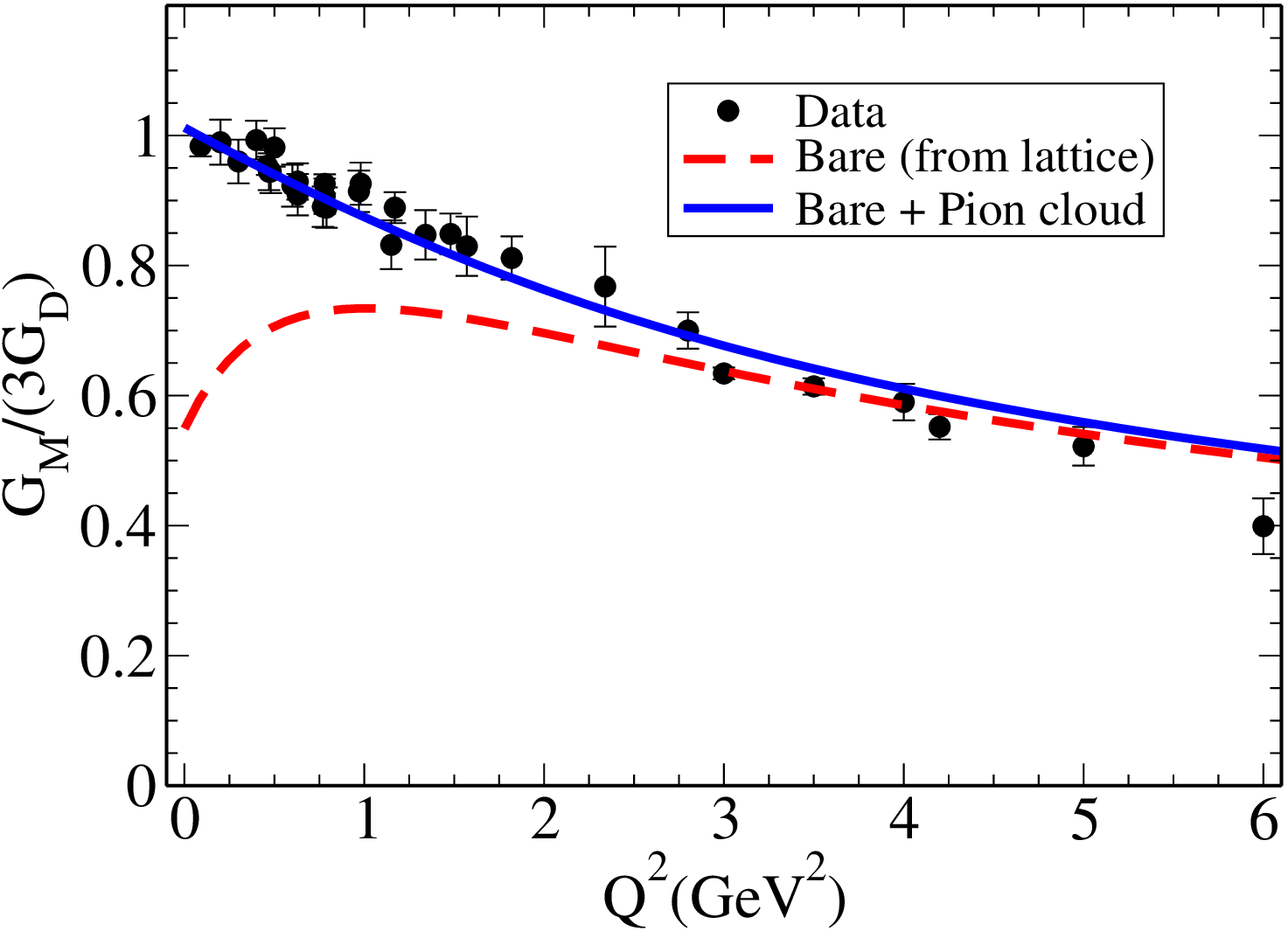} \hspace{.8cm}
\includegraphics[width=3.1in]{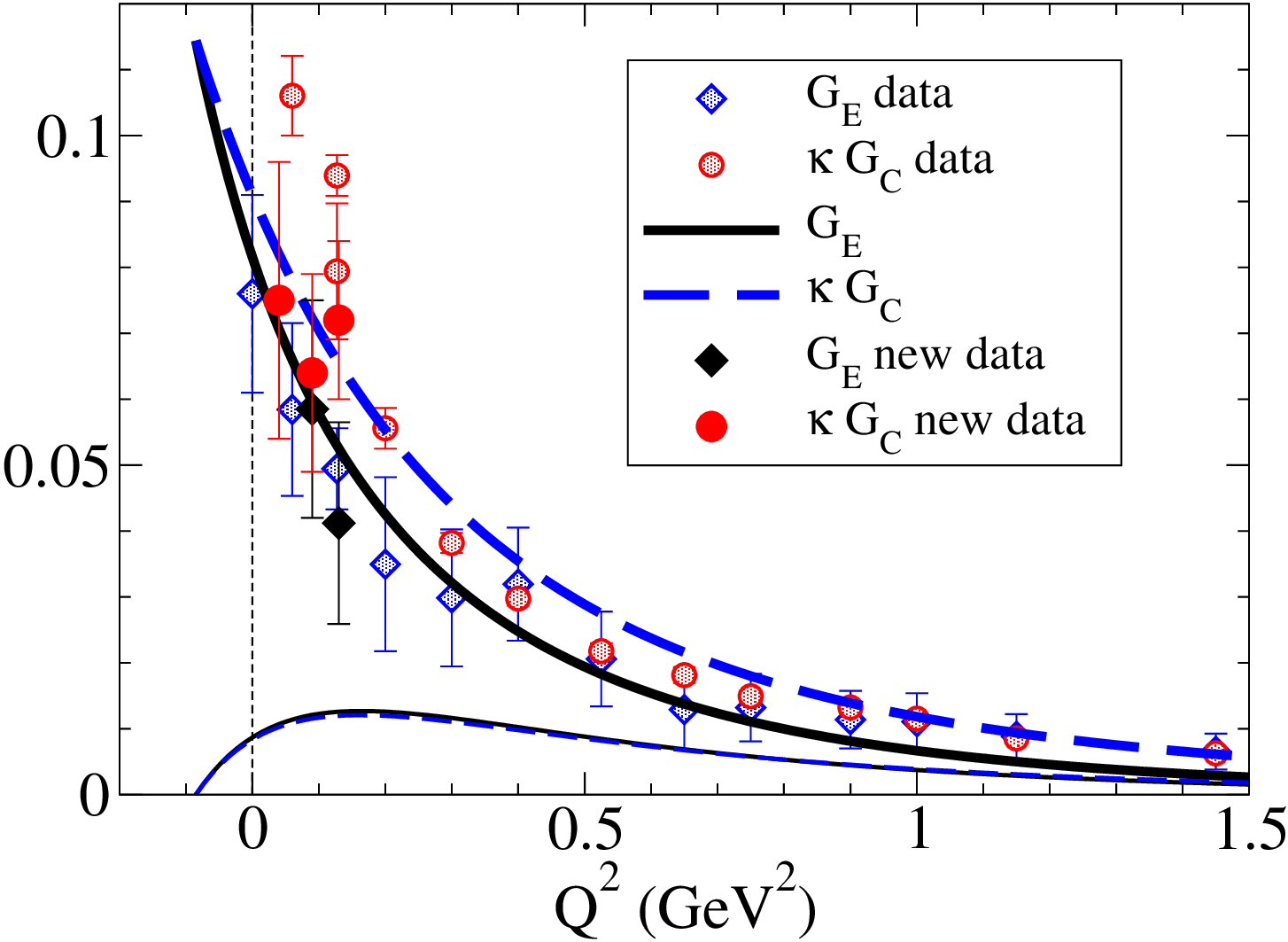}
}}
\caption{\footnotesize
$\gamma^\ast N \to \Delta$ transition form factors.
{\bf Left:} Magnetic dipole form factor normalized by the dipole form 
factor $G_D= 1/(1 + Q^2/\Lambda_D^2)^2$, where $\Lambda_D^2 = 0.71$ GeV$^2$.
{\bf Right:} Quadrupole form factors $G_E$ and $G_C$, 
$\kappa= \frac{M_\Delta -M_N}{2 M_\Delta}$.
Data from Refs.~\cite{MokeevDatabase} (open circles and diamonds)
and \cite{Blomberg16a} (solid circles and diamonds).
The thin-lines represent the valence quark contributions 
for $G_E$ (solid line) and for $G_C$ (dashed-line).}
\label{figDelta}
\end{figure*}

The results 
for $N(1440)$  are present in the Fig.~\ref{figRoper} (solid-line) 
in comparison with the CLAS/JLab data~\cite{MokeevDatabase}.
In the figure, we can notice that the model describes very well 
the $Q^2 > 2$ GeV$^2$ data, corroborating the idea 
that $N(1440)$ is in fact the radial excitation of the nucleon.
The deviations at small $Q^2$ can be interpreted as 
a consequence of the meson cloud effects, omitted in the present model,
or as a consequence of the 
approximated form considered for the orthogonality condition,
since the orthogonality was imposed only in the first order of $(M_R-M)^2$. 
More details can be found in Appendix B from Ref.~\cite{Roper}.
In a recent work, the valence quark contributions 
to the $\gamma^\ast N \to N(1440)$ form factors have 
been calculated within a holographic model~\cite{Roper-AdS,Roper-AdS2}.
The results are also presented in Fig.~\ref{figRoper} by the red band. 
Those results suggest that holographic methods 
can be used to estimate the valence quark effects, even at small $Q^2$,
and that, in the case of the Pauli form factor, 
the effects of the meson cloud may be small~\cite{Roper-AdS}.

The covariant spectator quark have been also applied to 
the study of the $\Delta(1232)\frac{3}{2}^+$ resonance.
The wave function associated with the $\Delta(1232)$ 
can be written as a combination of three 
angular momentum states 
for the quark-diquark system: 
a $S$ state and  two different $D$ states,
labeled as  $D1$ and $D3$
(core spin 1/2 and 3/2, respectively)~\cite{NDeltaD,LatticeD,DeltaDFF}.
The spin-isospin structure is determined 
by the respective symmetries.
For the radial wave functions, $\psi_\Delta^{S,D1,D3}$,
one needs, however, to use some asantz.
In this case we cannot relate 
the $\Delta$ radial wave functions with the nucleon radial wave function.
There are two main reasons for that:
i) the nucleon and the $\Delta$ are based on very different 
spin-isospin states;
ii) in the case of the $\Delta$, we cannot parametrize 
the radial wave function using directly
the elastic data ($\gamma^\ast \Delta \to \Delta$), 
neither by the $\gamma^\ast N \to \Delta$ data,
since there are evidences that the data are strongly 
contaminated by meson cloud effects~\cite{NDelta,NDeltaD,NDeltaTL,JDiaz07}.
One uses, therefore, lattice QCD data~\cite{Alexandrou} 
to determine  $\Delta$ radial wave functions $\psi_\Delta^{S,D1,D3}$.

The extension of the model to the lattice QCD regime 
takes advantage of 
two properties of the model:
the representation of the quark form factors $f_{i\pm}$
in terms of the vector meson dominance mechanism
(implicit dependence on the rho mass $m_\rho$):
$f_{i\pm}(Q^2) \equiv f_{i\pm} (Q^2;m_\rho,M)$,
and the representation of the radial wave 
functions in terms of 
the baryon mass, $M_B$.
In the lattice QCD regime, we can then replace the dependence 
of the masses ($m_\rho$, $M$ and $M_B$) in the 
physical regime by the lattice QCD masses. 
More details can be found in Refs.~\cite{Omega,LatticeD,Lattice}.
Once determined the radial wave functions by 
the lattice QCD data, the model is extrapolated 
to the physical regime, and used to calculate 
the valence quark contributions for each form factor.
The results are presented in Fig.~\ref{figDelta}, for 
the magnetic dipole $G_M$ (dashed-line) and 
for the electric $G_E$ and Coulomb $G_C$ quadrupole form factors
(thin-lines near the horizontal axis).

For the magnetic dipole form factor, $G_M$,
in the left panel of Fig.~\ref{figDelta},
one can observe that the valence quark contribution (Bare), 
estimated with the assistance of the lattice QCD data  (dashed-line),
underestimate the data below $Q^2=2$ GeV$^2$.
Only for larger values of $Q^2$ there is 
the convergence with the full result (solid-line)~\cite{LatticeD}.
In the present study the full result (Bare + Pion cloud) 
is obtained using a phenomenological parametrization 
to the pion cloud component 
$G_M^\pi \propto \left( \frac{\Lambda_\pi^2}{\Lambda_\pi^2 + Q^2}\right)^2$,
with an adjustable strength coefficient and 
a cutoff $\Lambda_\pi$~\cite{NDeltaD,LatticeD}.
The estimates for the bare and meson cloud contribution
are in good agreement with the estimates 
from the  EBAC/JLab dynamical coupled-channel model 
for the baryon-meson reactions~\cite{LatticeD,JDiaz07}. 

Our estimates for the electric and Coulomb quadrupole 
form factors are presented in the left panel of Fig.~\ref{figDelta}.
For convenience, we present the results for $G_C$ 
multiplied by the factor $\kappa= \frac{M_\Delta -M_N}{2 M_\Delta}$,
where $M_\Delta$ is the $\Delta(1232)$ mass.
In the figure, one can notice the very good agreement 
between the final results (tick-lines), which 
include the bare and the pion cloud contributions, 
and the overall data~\cite{Delta-Qudrupole,RSM-Siegert}.
Only at low $Q^2$, there are some discrepancy 
with the $G_C$ data, which has been interpreted as 
a  consequence of errors in the analysis of the data.
The old low-$Q^2$ data have replaced by a more recent and 
reliable analysis~\cite{RSM-Siegert,Blomberg16a}.
The most recent results are represented by the 
solid circles and diamonds.
For the good agreement between theory and data
contribute the combination of the small valence contributions 
($\approx 10\%$, thin-lines) and the pion cloud contributions.
The pion cloud contributions are estimated 
by large $N_c$ parameter-free 
relations~\cite{RSM-Siegert,Buchmann04,Pascalutsa07a}. 
Since the valence quark contribution is fixed by the lattice QCD data,
the final results are true predictions~\cite{RSM-Siegert}.
The convergence between the results for $G_E$ and $\kappa G_C$ 
at the pseudothreshold, when $Q^2=-(M_\Delta -M_N)^2$ 
is a consequence of Siegert's theorem~\cite{RSM-Siegert,Delta-Siegert}.

\section*{SUMMARY AND OUTLOOK}

The covariant spectator quark model, successful 
in the description of the nucleon elastic 
form factors revealed by the JLab 
polarization transfer experiments, 
has been extended to the calculation 
of transition form factors associated 
with several light nucleon resonances $N^\ast$.
Combining the electromagnetic structure of the 
constituent quarks, calibrated by the nucleon data, 
with appropriate asantz for the radial wave functions 
of the resonances, we are able to estimate the 
transition for factors for the resonances 
$N(1440)\frac{1}{2}^+$,
$N(1535)\frac{1}{2}^-$ and
$N(1520)\frac{3}{2}^-$.
The estimates are based on 
the parametrization of the nucleon radial wave function 
(nucleon shape) with no adjustable parameters.
The calculations take into account, exclusively,
the effect of the valence quarks, and are 
in good agreement with the data for momentum transfer $Q^2 > 2$ GeV$^2$, 
with a few exceptions.
The exceptions may be explained by meson cloud effects.
All the estimates are based on 
the parametrization of the nucleon structure,
and can be tested in a near future 
for large transfer momentum in the 
Jefferson Lab -- 12 GeV upgrade.    
The model has been also extended to the $\Delta(1232)\frac{3}{2}^+$, 
with the assistance of the lattice QCD, to estimate 
the radial wave functions of the $\Delta(1232)$ valence quark core.
The model estimates are in agreement with the 
empirical data when theoretical and phenomenological 
parametrizations of the pion cloud are taken into account.

\vspace{-.3cm}
\begin{acknowledgments}
\vspace{-.2cm}
This work was supported by the Funda\c{c}\~ao de Amparo \`a 
Pesquisa do Estado de S\~ao Paulo (FAPESP):
project no.~2017/02684-5, grant no.~2017/17020-BCO-JP. 
\end{acknowledgments}



\section{Details of the model}

The radial wave function of the nucleon, which encode 
the dependence on the nucleon ($P$) and diquark ($k$) momentum,
can be represented  in terms of the dimensionless variable   
\ba
\chi = \frac{(M_N-m_s)^2 -(P-k)^2}{M m_s},
\ea
where $M_N$ is the mass of the nucleon and 
$m_s$ is the mass of the diquark.
This particular dependence is possible because
in the covariant spectator quark model
the nucleon and the diquark are both on mass shell 
(more details can be found in Ref.~[3]).

The explicit form of the radial wave function is
\ba
\psi_N(\chi) = \frac{N_0}{m_s(\beta_1 + \chi)(\beta_2 + \chi)},
\ea 
where $N_0$ is a normalization constant and 
$\beta_i$ are dimensionless parameters in units $M_N m_s$,
which can be converted into square momentum scales.

The representation of the quark form factors
is motivated by the vector dominance model
\ba
& &
f_{1\pm} (Q^2) =
\lambda + (1- \lambda) \frac{m_v^2}{m_v^2 + Q^2} + 
c_\pm \frac{M_h^2 Q^2}{(M_h^2 + Q^2)^2}, \nonumber \\
& &
f_{2\pm} (Q^2) =
\kappa_\pm \left[
d_\pm
\frac{m_v^2}{m_v^2 + Q^2} + (1- d_\pm)
 \frac{M_h^2 }{M_h^2 + Q^2} \right], \nonumber \\
& &
\ea
where $\kappa_\pm$ are the quark isoscalar/isovector anomalous 
magnetic moments, $m_v$ is a vector meson mass and $M_h$ 
is fixed heavy mass parameter (short range scale).
In the model those are fixed as $m_v = m_\rho \simeq m_\omega$.
The quark anomalous magnetic moments
are determined by the proton and neutron 
magnetic moments ($Q^2=0$).  
$\lambda$ is a parameter fixed in the study of the 
deep inelastic scattering.
The remaining parameters $c_\pm$ and 
$d_\pm$ are determined by the phenomenology, 
more specifically by the fit to 
the nucleon elastic form factors.

The minimal model (model II from Ref.~[2]) 
fixes $M_h= 2 M_N$ and $d_+ = d_-$ leaving 
only 3 free parameters for the quark form factors
($c_+$, $c_-$ and $d_+$).

\end{document}